\documentclass{sig-alternate}
\usepackage{mathptmx} %
\usepackage{fancyhdr}
\usepackage[normalem]{ulem}
\usepackage[hyphens]{url}
\usepackage[sort,nocompress]{cite}
\usepackage[final]{microtype}
\usepackage[keeplastbox]{flushend}
\usepackage[bookmarks=true,breaklinks=true,letterpaper=true,colorlinks,citecolor=blue,linkcolor=blue,urlcolor=blue]{hyperref}
\fancypagestyle{firstpage}{
  \fancyhf{}
  
}
\usepackage{amsmath,mathtools,amssymb}
\usepackage{titlesec}
\usepackage{multirow}
\usepackage{tikz}
\usepackage{adjustbox}
\usepackage[font=small,labelfont=bf,textfont=bf]{caption}
\titlespacing*{\section}
{0pt}{\parsep}{\parsep}
\titlespacing*{\subsection}
{0pt}{\parsep}{\parsep}
\usepackage[ruled,vlined,linesnumbered]{algorithm2e}

\newcommand{\lcell}[2]{\parbox{#1}{\vspace{2pt}\raggedright #2 \\\vspace{-8pt}\hphantom{}}}

\def\frameworkName{CECQ}
\def\titleName {Enabling Full-Stack Quantum Computing with Changeable Error-Corrected Qubits}
\author{
Anbang Wu\\
anbang@ucsb.edu \\
UC, Santa Barbara
\and
Keyi Yin\\
keyi@ucsb.edu \\
UC, Santa Barbara
\and
Andrew W. Cross\\
awcross@us.ibm.com \\
IBM T.J Watson Research Center
\and
Ang Li\\
ang.li@pnnl.gov \\
Pacific Northwest National Laboratory
\and
Yufei Ding\\
yufeiding@cs.ucsb.edu \\
UC, Santa Barbara
}
\title{\titleName} 

\begin{document}
\maketitle
\thispagestyle{firstpage}
\pagestyle{plain}
\begin{abstract}
  Executing quantum applications with quantum error correction (QEC) faces the gate non-universality problem imposed by the Eastin-Knill theorem. As one resource-time-efficient solution, code switching changes the encoding of logical qubits to implement universal logical gates.
  Unfortunately, it is still unclear how to perform full-stack fault-tolerant quantum computing (FTQC) based on the changeable logical qubit. Specifically, three critical problems remain unsolved: a) how to implement the dynamic logical qubit on hardware; b) how to determine the appropriate timing for logical qubit varying; c) how to improve the overall system performance for programs of different features. To overcome those design problems, We propose \frameworkName, to explore the large design space for FTQC based on changeable logical qubits. Experiments on various quantum programs demonstrate the effectiveness of \frameworkName{}.
\end{abstract}

\section{Introduction}\label{sect:intro}
Quantum computing suffers from device noise which greatly limits the problem size a quantum device can address with a low failure rate~\cite{preskill2018quantum}. Quantum error correction (QEC) codes are widely studied to mitigate quantum noises and enable fault-tolerant quantum computing (FTQC)~\cite{nielsen2002quantum}. QEC encodes a high-fidelity logical qubit with a group of unreliable physical qubits (named data qubits) and corrects potential errors on the logical qubit based on error information extracted from data qubits. Experiments demonstrate the reliability of quantum architectures based on QEC and the enablement of QEC on realistic hardware has witnessed a series of breakthroughs recently~\cite{surfstitch,Chamberland2019TriangularCC,Lao2020FaulttolerantQE,Chamberland2020TopologicalAS,sundaresan2022matching,google2023suppressing,ryan2022implementing,krinner2022realizing,egan2020fault}.
Executing quantum applications upon any QEC code faces the unique problem of gate nontransversality imposed by the Eastin-Knill theorem~\cite{eastin2009restrictions}, which indicates that not all logical gates in a \textbf{universal} gate set (e.g., Clifford+T~\cite{nielsen2002quantum}) can be implemented transversely, i.e., by applying physical gates where each acts on exactly one data qubit of the logical qubit. For instance, the logical T gate $T_L$ is not a transverse gate of the Steane code~\cite{steane1996multiple}. 
To build a universal gate set to accommodate any programs, various protocols, e.g., magic distillation and code switching, are proposed to implement non-transverse logical gates. Among them, code switching stands out for its potentially smaller resource and time overhead~\cite{CostofUniversality} on future less error-prone quantum hardware. The code switching protocol allows the implementation of non-transverse logical gates by encoding the logical qubit in different QEC codes along the time dimension. For instance, to implement $T_L$ on a logical qubit of the Steane code, one potential code switching is to transform the logical qubit to the Reed-Muller (RM) code which supports transverse $T_L$. After $T_L$ is executed, we can then use another code switching to transform the logical qubit back into the Steane code.
Existing efforts on code switching only consider implementing specific logical gates, e.g., $T_L$. It is unclear how we can implement QEC with code switching (in short, QEC-CS) on quantum hardware. Indeed, we discover a unique and unexplored design space for implementing the QEC-CS architecture. 
\textbf{Firstly}, we claim that each implemented logical qubit of QEC-CS should support a changeable data qubit layout to allow two QEC code types. When the logical qubit is switched from one code to the other code, the underlying data qubits and error detection circuits will be changed accordingly. 
\textbf{Secondly}, we observe that the data qubit layout optimization of the QEC-CS logical qubit involves more factors. Compared to the single-typed QEC architecture~\cite{javadi2017optimized}, besides error detection circuits, the layout of the QEC-CS logical qubit also affects the reliability of code switching. 
\textbf{Finally}, considering the importance of logical gates, the design of the QEC-CS architecture should balance the performance of error detection, code switching, and logical gates simultaneously. We should not only optimize the data qubit layout of one logical qubit but also the placement of multiple QEC-CS logical qubits.
Overall, the architectural design space of QEC-CS identified by our paper is far beyond the scope of existing works that map QEC to hardware~\cite{surfstitch,Chamberland2019TriangularCC,Lao2020FaulttolerantQE,Chamberland2020TopologicalAS}. Existing works focus on the implementation of one single-typed logical qubit and their optimization goal is only to reduce the error detection overhead of the single-typed logical qubit.
Besides the lack of efforts on architectural design, we further observe that existing software support for FTQC~\cite{javadi2017optimized,Hua2021AutoBraidAF,Paler2019SurfBraidAC,Ding2018MagicStateFU} misses unique compiler optimization opportunities in the QEC-CS architecture. Conventionally, to adapt a general program to QEC architectures, existing QEC compilers~\cite{javadi2017optimized,Hua2021AutoBraidAF} decompose the program into the Clifford+T basis, map each program qubit to a logical qubit, and then execute the resulting logical circuit in a similar way to treating quantum circuits without QEC. We argue that this simple compilation support does not unveil the computational potential of QEC-CS. Code switching, as the most important enabler of the QEC-CS architecture, is an expensive QEC operation. Code switching requires a series of noisy and time-consuming physical operations between data qubits in one logical qubit~\cite{CostofUniversality}. Reducing the utilization of code switching is critical for minimizing the space-time overhead and error rate of quantum programs on the QEC-CS architecture. Unfortunately, code switching still remains unoptimized in existing QEC compilers. Overall, to efficiently utilize the QEC-CS architecture, it is critical to optimize the code switching usage, rather than sticking to the conventional QEC compilers. 
In this paper, we propose the first full-stack framework, named~\frameworkName, to provide architecture and software support for FTQC based on QEC-CS. Firstly, we present a comprehensive architectural design for the QEC-CS logical qubit, accounting for the data qubit layout changeability and the interplay of error detection, code switching, and logical quantum gates.
We observe that those three QEC operations have different impacts on the architecture reliability and space overhead. Their hardware implementation may even conflict with each other. To address those problems and enable effective exploration of the large architecture design space, we propose using the profiling data (e.g., impact on logical operation error rates, consumed physical qubits) and structural information (e.g., data qubit locations in an error detection circuit) of those QEC operations to guide the search priority and circumvent implementation conflicts. Experiments demonstrate the effectiveness of our architectural design over the large search space.
Secondly, we present a compiler design that supports the efficient execution of quantum programs on the QEC-CS architecture. On the one hand, we observe that a pair of code switching operations can be used to execute more than the logical T gate depending on the program context, saving program latency and improving fidelity. On the other hand, we observe that executing too many logical gates between a pair of code switching operations may instead hurt the program fidelity considering the different error correction capabilities of the two QEC codes being switched. Our compiler achieves a good balance for this unique trade-off in the QEC-CS architecture and far surpasses existing QEC compilers in reducing the space-time overhead of quantum programs, as demonstrated by our experiments.
Finally, we present a co-design of QEC-CS architecture and compiler to further promote the computational potential of QEC-CS. We observe that, for compiled quantum programs of distinguished features, regarding different optimization metrics, it is better to use different connectivity between logical qubits and place QEC-CS logical qubits accordingly.
This is the first co-design study of FTQC since existing works~\cite{surfstitch,Chamberland2019TriangularCC,Lao2020FaulttolerantQE,Chamberland2020TopologicalAS} only consider the simulation of one logical qubit. Our evaluation shows that the co-design can further improve the performance of specific quantum programs in particular metrics, e.g., the space-time overhead and the program fidelity.

\section{Background} \label{sect: bg}
In this section, we would introduce three critical protocols for FTQC: QEC, logical quantum gates, and the specific protocol that helps implement logical gates, i.e., code switching in the paper. We would also introduce the necessary background knowledge to implement QEC codes on quantum hardware.
\subsection{Fault-tolerant Quantum Computing}
\begin{figure}[t]
    \centering
    \includegraphics[width=0.46\textwidth]{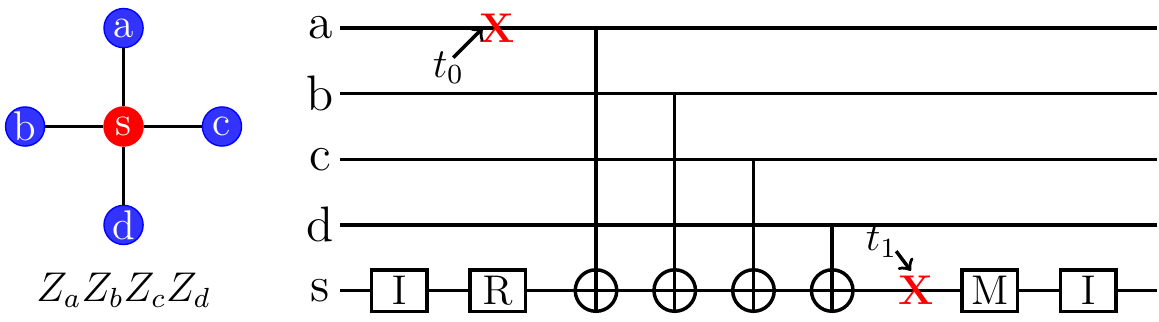}\hphantom{-5pt}
    \caption{Error detection circuits to detect Pauli X errors from data qubits $a, b, c, d$. R, I, and M are reset, identity gate, and measurement respectively. Red X denotes Pauli X error.}
    \label{fig:stabilizercirc}\hphantom{-8pt}
\end{figure}

The \textbf{first} enabler of FTQC is the QEC code, which is proposed by theorists to eliminate quantum hardware noise. QEC codes encode a logical qubit with many noisy physical qubits which are named data qubits in the QEC context. With this redundancy, QEC codes can detect potential errors in data qubits of a logical qubit by using error detection circuits. 
Figure~\ref{fig:stabilizercirc} shows the circuit used to measure the weight-four stabilizer operator $Z_aZ_bZ_cZ_d$, to detect Pauli X errors that occurred on four data qubits $\{a, b, c, d\}$ (blue dots). Likewise, the stabilizer operator $X_aX_bX_cX_d$ detects Pauli Z errors on $\{a, b, c, d\}$.
In error detection circuits, the error information of data qubits is aggregated to the parity qubit $s$ (red dots) by physical CX gates. For instance, in Figure~\ref{fig:stabilizercirc}, the X error on qubit $a$ at time $t_0$ would lead to the X error on $s$ at $t_1$ and flip the result of the following measurement operation. 
After error detection, the measurement result will be fed into a QEC decoder to identify the erroneous qubit and suggest the best error correction operations. 
\begin{figure*}[t]
    \centering
    \includegraphics[width=0.995\textwidth]{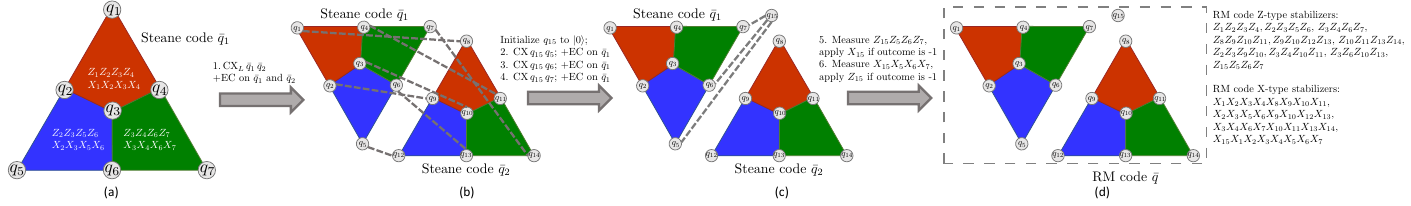}\hphantom{-10pt}
    \caption{Code conversion from the Steane code to the RM code~\cite{CostofUniversality}. Dotted lines denote physical CX gates. EC means error detection and correction.}
    \label{fig:codeconv}
\end{figure*}

The \textbf{second} enabler of FTQC is fault-tolerant logical gates. For each underlying physical gate of the fault-tolerant logical gate, if it induces physical errors on data qubits, the induced errors should be detectable and correctable in the current QEC code. 
Fortunately, transverse logical gates are naturally fault-tolerant. Being transverse implies it is constructed by posing exactly one physical gate on each data qubit of the logical qubit. For instance, on a logical qubit $\bar{q}$ of the Steane code (in short, Steane logical qubit) with data qubits $q_1,q_2,\cdots,q_7$, the logical H gate ($H_L$) is transversal and is defined by $H_L\bar{q} = \otimes_{i=1}^7 Hq_i$. Logical CX ($CX_L$) is also transverse and is defined by $CX_L\bar{q}_0\bar{q}_1 = \otimes_{i=1}^7 CXq_{0i}q_{1i}$.
For transverse logical gates, one physical gate error at most induces one data qubit error per logical qubit, which is correctable for any QEC codes of code distance $\ge$3 (e.g., Steane code) by definition (of code distance~\cite{nielsen2002quantum}). Unfortunately, not all logical gates of one QEC code have a transverse implementation, according to the Eastin-Knill theorem~\cite{eastin2009restrictions}. For instance, the logical T gate is transversal in the 15-qubit Reed-Muller (RM) code (defined as $T_L \bar{q} = \otimes_{i=1}^{15} T^\dagger q_i$) but is non-transverse in the Steane code. Likewise, the logical H gate is non-transverse in the RM code. To achieve universal FTQC with the Clifford+T basis, various schemes (e.g., magic distillation~\cite{bravyi2005universal} and code switching~\cite{codeconvsteane}) are proposed to provide fault-tolerant implementations for non-transverse logical gates. Among these schemes, code switching stands out for its potentially smaller resource-time overhead~\cite{CostofUniversality} on future less erroneous  hardware.
For FTQC based on QEC-CS, the \textbf{third} enabler is the code switching protocol.
To implement non-transverse logical gate fault-tolerantly, code switching encodes the logical qubit in different QEC codes along the time dimension, e.g., switching between the Steane and the RM code, which is widely studied in existing works~\cite{codeconvsteane, CostofUniversality}. Commonly, to implement the logical T gate on a Steane logical qubit, we would transform the logical qubit to the RM code. After the logical T gate is transversely executed, we would then use one more code switching operation to transform the logical qubit back into the Steane code. 
Figure~\ref{fig:codeconv} shows the code switching from the Steane code to the RM code, which converts the Steane logical qubit state into an RM logical qubit state. In the figure, one RM logical qubit contains two Steane logical qubits, with $q_{15}$  used to establish the connection between the two Steane logical qubits.
The whole code switching process in Figure~\ref{fig:codeconv} contains one logical qubit between two Steane logical qubits, three physical CX gates, three Steane error detection rounds, and one RM error detection round. The code switching from the RM code to the Steane code is just the reverse of the process in Figure~\ref{fig:codeconv}.

\subsection{Enforcing QEC on Hardware}
\begin{figure}[t]
    \centering
    \includegraphics[width=0.46\textwidth]{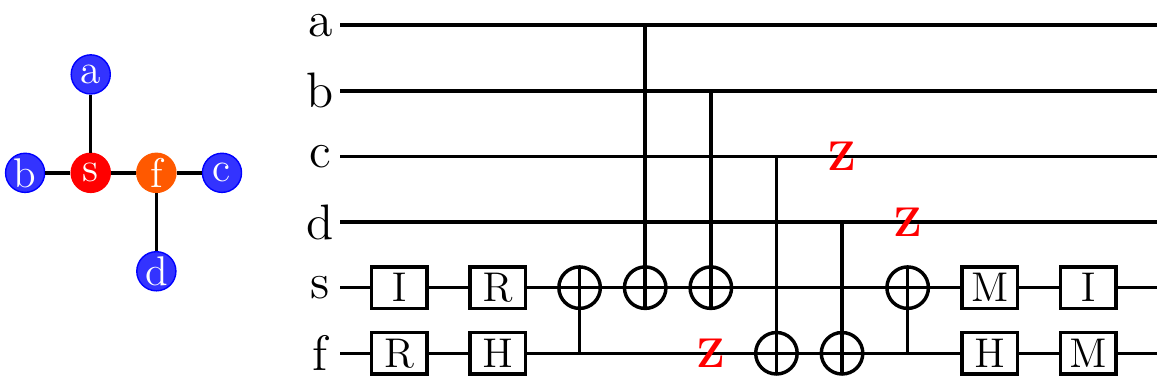}\hphantom{-5pt}
    \caption{The flag-bridge circuit to detect Pauli X errors from data qubits $a, b, c, d$. R, H, I, and M are reset, Hadamard gate, identity gate, and measurement respectively. Red Z indicates Pauli Z error.}
    \label{fig:flagcirc}\hphantom{-2pt}
\end{figure}

In many quantum devices (e.g., superconducting~\cite{Qiskit} and neutral atom hardware~\cite{henriet2020quantum}), the connectivity between physical logical qubits may be limited. In such a case, the physical CX gate in error detection circuits, logical CX, and code switching may be on non-neighboring physical qubits.
To consider the topology constraints, in case the parity qubit is not directly linked to data qubits, we use the widely-used flag-bridge circuit~\cite{Lao2020FaulttolerantQE,surfstitch}(see Figure~\ref{fig:flagcirc}) rather than the SWAP approach for efficiency and high error detection accuracy~\cite{surfstitch}. In Figure~\ref{fig:flagcirc}, data qubits $\{c, d\}$ are not in the neighborhood of the parity qubit $s$, which means we cannot apply CX gates between those data qubits and $s$. The flag qubits $f$ (orange dots) neighboring $s$ would then be used to help gather error information from those data qubits. The parity information collected by $f$ would be propagated to $s$ by physical CX gates between them. The flag-bridge circuit guarantees the fault tolerance of error detection. For example in Figure~\ref{fig:flagcirc}, the Z error on $f$ may lead to correlated Z errors on data qubit $\{c,d\}$. Fortunately, the Z error on $f$ would flip measurement result on $f$, making the correlated Z errors on $\{c,d\}$ detectable and correctable. 

\begin{figure}[t]
    \centering
\hphantom{}\hspace{-10pt}\includegraphics[width=0.48\textwidth]{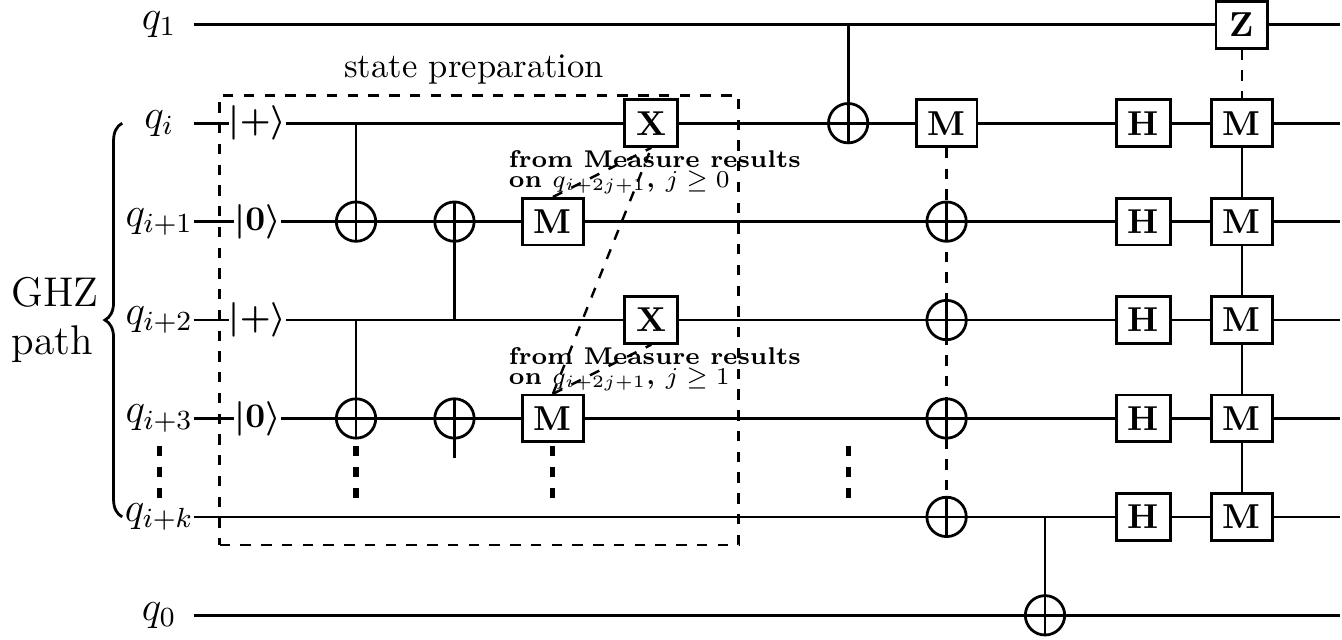}\hphantom{-5pt}
    \caption{Remote CX between $q_1$ and $q_0$ using the GHZ state.}
    \label{fig:ghz}\hphantom{-8pt}
\end{figure}

For logical CX and code switching, to perform physical CX gates between data qubits that are not physically coupled, we use a GHZ-state based approach (see Figure~\ref{fig:ghz}) rather than SWAP for low cost, as practiced widely by existing works~\cite{gottesman1999quantum}. In Figure~\ref{fig:ghz}, we build a GHZ state using $\{q_i, \cdots, q_{i+k}\}$ and the remote CX gate between $q_1$ and $q_0$ is implemented by using the prepared GHZ state.
For simplicity, we say that qubits in the state preparation part of Figure~\ref{fig:ghz} (i.e., $q_i,\cdots,q_{i+k}$) form a \textbf{GHZ path} to connect two data qubits.
\section{Design Considerations}\label{sec:design}
In this paper, our major objective is to improve the overall fidelity as well as reduce the space-time overhead of running quantum programs on the QEC-CS architecture by orchestrating the architecture and software design. In this section, we highlight the considerations and observations to achieve our design objectives. 
\subsection{Problem Setting}
In our architecture design for QEC-CS, we assume that the underlying quantum hardware has enough qubits to accommodate logical qubits of the QEC code, which is a common assumption for FTQC. We explicitly study the QEC-CS architecture design with code switching between the Steane  and RM code, but our design is not limited to them and can be extended to general 2D color code and 3D stabilizer color code. Further, we use the flag-bridge circuit in Figure~\ref{fig:flagcirc} for error detection and the GHZ-state-based method in Figure~\ref{fig:ghz} for the CX gate between non-neighboring physical qubits.

\subsection{Architectural design for QEC-CS}

Generally, to enforce QEC support on quantum hardware, we need to map data qubits and error detection circuits of logical qubits to the hardware~\cite{surfstitch}. 
However, for the QEC-CS logical qubit design, we should further support dynamic code switching between two QEC codes. Overall, when placing data qubits of a QEC-CS logical qubit on hardware, we expect the overhead (e.g., involved qubit/gate count, latency) of executing error detection, logical CX, and code switching on data qubits to be as small as possible. A QEC-CS logical qubit with lower overhead is more reliable because using more qubits and more gates would induce more error locations~\cite{nielsen2002quantum}.
\begin{figure}[h]
    \centering
\includegraphics[width=0.49\textwidth]{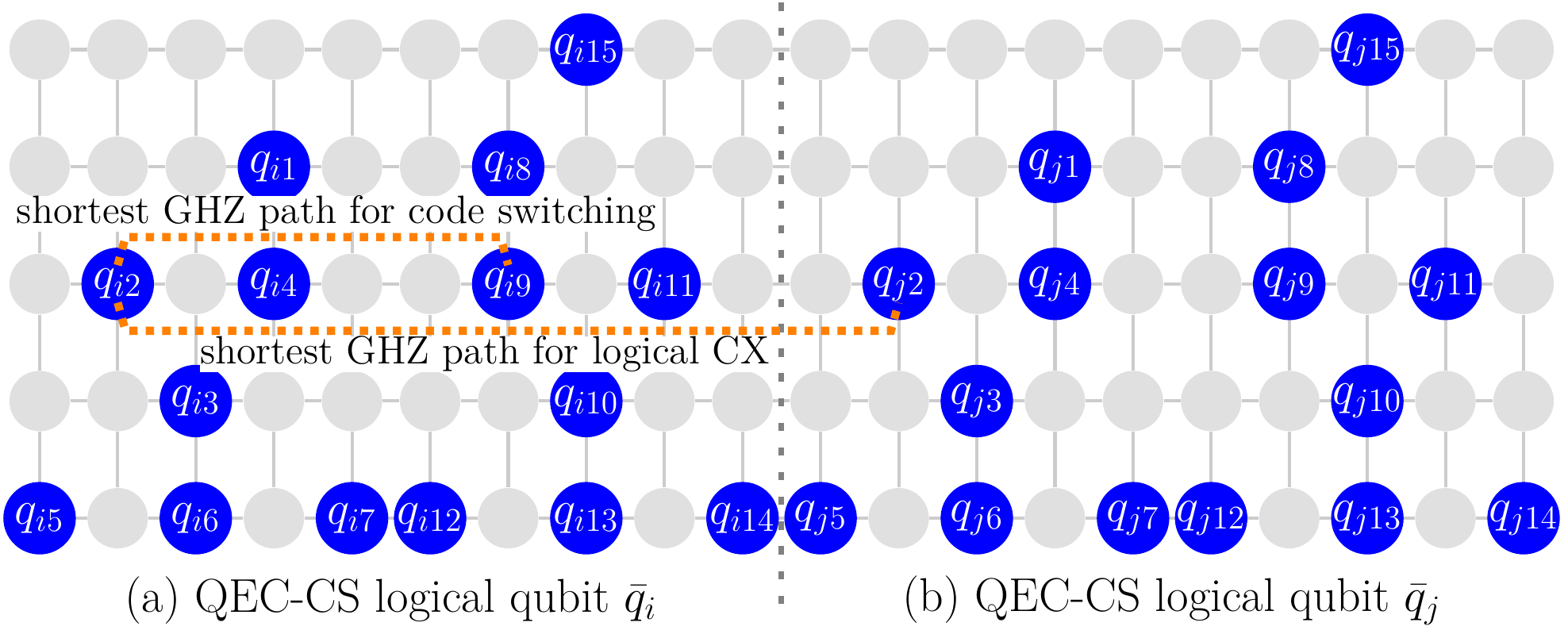}\hphantom{-8pt}
    \caption{An example data qubit layout for QEC-CS logical qubits, which only minimizes the error detection cost. }
    \label{fig:mot1}\hphantom{-5pt}
\end{figure}
Unfortunately, the overhead optimization of error detection differs from the optimization of code switching, and logical gates. The error detection circuit requires data qubits to be close to the parity qubit as much as possible so that we can use fewer qubit resources (e.g., flag qubits) to gather error information from data qubits that are not in the neighborhood of the parity qubit. 
However, placing data qubits too close would hinder the implementation of the logical CX gate as well as code switching. 
Figure~\ref{fig:mot1} shows an example of the data qubit layout that tries to minimize the qubit resource overhead of error detection circuits. 
Below both code switching and logical CX, we need to execute many physical CX gates between data qubits that are often not neighbors. For example, for code switching in QEC-CS logical qubit $\bar{q}_i$ in Figure~\ref{fig:mot1}(a), we need to execute a physical CX gate between $q_{i2}$ and $q_{i9}$. To perform the CX gate on two non-neighboring data qubits, we need a GHZ path to connect them, as shown in Figure~\ref{fig:mot1}.
As we can see, for code switching, the shortest GHZ path connecting data qubit $q_{i2}$ and $q_{i9}$ is blocked by $q_{i4}$; for logical CX, the shortest GHZ path connecting $q_{i2}$ and $q_{j2}$ is also blocked by both $q_{i4}$. This means placing data qubits too close may increase the overhead (e.g., the length of GHZ paths) of implementing code switching and logical CX.

Similarly, the overhead optimization of code switching differs from the optimization of the logical CX gate. For code switching, the logical CX is between two Steane logical qubits that together form an RM logical qubit. All data qubits and logical gates involved in the code switching are within one logical qubit of QEC-CS. To reduce the space overhead of code switching, we would make data qubits involved in the logical CX between two `interior' Steane logical qubits as close as possible to reduce the length of GHZ paths. 
However, this optimization may increase the overhead of the logical CX between two QEC-CS logical qubits, similar to the discussion above. For example, in Figure~\ref{fig:mot1}, the shortest GHZ path between $q_{i2}$ and $q_{j2}$ is blocked by $q_{i9}$, which is involved in a physical CX gate of the code switching.
Overall, it remains unclear how to design the interior layout of a logical qubit so that we can guarantee low overhead and good reliability for error detection circuits, code switching, and logical gates, simultaneously.
To make good choices for the QEC-CS logical qubit design, we believe it is critical to gather quantitative data showing the impacts of the three types of operations on the reliability of the architecture based on QEC-CS. 

\begin{table}[h]
\centering
\caption{Pseudo-threshold of the Steane logical CX gate for different \# flag qubit, and GHZ path length (averaged for all pairs of data qubits).}\label{table:scx}
\resizebox{0.46\textwidth}{!}{
\begin{tabular}{|p{2.8cm}|c|c|c|c|c|}
\hline
\hphantom{}\hspace{-9pt}\multirow{2}{*}{\begin{tikzpicture}\node at (0,1) {\hspace{3pt}\parbox{0.8cm}{\hphantom{-3.5pt}\# flag qubit}}; \draw (0.9,0.7) -- (0.5,1.45);\node at (1.7,1) {\parbox{1.4cm}{\hphantom{-3.5pt} GHZ path length}};\end{tikzpicture}}
 & \multirow{2}{*}{0} & \multirow{2}{*}{1} & \multirow{2}{*}{2} & \multirow{2}{*}{3} & \multirow{2}{*}{4} \\ 
 &  &  &  &  &  \\ \hline
\parbox{2cm}{\hspace{30pt} 1} & 0.0034 & 0.0022 & 0.0016 & 0.0012 & 0.0009 \\ \hline
\parbox{2cm}{\hspace{30pt} 2} & 0.0017 & 0.0013 & 0.0010 & 0.0008 & 0.0006 \\ \hline
\parbox{2cm}{\hspace{30pt} 3} & \multicolumn{1}{l|}{0.0010} & \multicolumn{1}{l|}{0.00081} & \multicolumn{1}{l|}{0.00066} & \multicolumn{1}{l|}{0.00054} & \multicolumn{1}{l|}{0.00046} \\ \hline
\end{tabular}
}\hphantom{-5pt}
\end{table}
Table~\ref{table:scx} shows the profiling data for the logical error rate of the Steane logical CX gate based on different design options for error detection and the logical CX gate. In Table~\ref{table:scx}, the device noise follows the circuit noise model~\cite{gidney2021stim}, and the decoder for the Steane code is the look-up table decoder~\cite{chamberland2018deep}.
We observe that it is more critical to reduce the overhead of error detection. One more flag qubit in the error detection circuit would decrease the pseudo-threshold of the logical CX more than having one more edge in the GHZ path between data qubits. This is because the error correction following the logical CX gate could largely mitigate the negative effect of longer GHZ paths. Further, increasing the overhead of error detection would also hurt the fidelity of other logical operations, e.g., code switching and single-logical-qubit gates.
Further, we observe that the logical CX gate should be optimized before code switching. This is because the logical CX gate may appear more frequently in  quantum programs. For example, quantum arithmetic circuits are based on the Toffoli gate. In the fully-connected qubit layout, the decomposed Toffoli gate has seven T gates and six CX gates~\cite{nielsen2002quantum}. Thus, a Toffoli gate at most requires fourteen code switching operations as each logical T gate requires two code switching operations.
However, for a connectivity-limited qubit layout, the overall CX count (i.e., 15) would surpass the amount of code switching required, as long as each CX gate requires 0.5 SWAP gates for routing. Note than this condition neglects the opportunity of reducing code switching operations by compiler optimizations.

\noindent\textbf{Observation 1}: a good data qubit layout for a QEC-CS logical qubit should first ensure the reliability of error detection, then improve the logical CX, and finally optimize overhead of code switching by fine-tuning the data qubit layout.

After the design of one QEC-CS logical qubit, the next step is to design the layout of multiple QEC-CS logical qubits for running quantum programs. 
There are two major design goals of placing multiple logical qubits. The first one is to reduce the latency and resource overhead (i.e., GHZ path length) of logical CX gates. The second one is to design the `logical' connectivity between logical qubits so that we can reduce the error rate/latency of quantum programs by reducing the (logical) SWAP gates caused by routing.
The placement of multiple QEC-CS logical qubits has a direct impact on the overhead of the logical CX gate. On the one hand, if we leave sufficient space between two logical qubit blocks, there would be enough physical qubits for parallel GHZ paths, thus reducing the latency of the logical CX gate. On the other hand, placing logical qubits farther away would increase the GHZ path length between two logical qubits thus decreasing the logical CX fidelity. Moreover, having more ancillary physical qubits between logical qubits would increase the qubit overhead of the overall logical qubit layout, which may instead cause the increase of the space-time overhead of quantum programs. We argue that it is more advantageous to place logical qubits close as the latency reduction by the sparse logical qubit layout may not cancel out the physical qubit overhead, in terms of the space-time overhead of quantum programs.
The placement of multiple QEC-CS logical qubits also directly affects the `logical' connectivity between logical qubits. The relative location of logical qubits affects whether it is worth enabling more direct logical CX gates for better connectivity.
On the one hand, higher logical connectivity reduces the SWAP gate for routing thus improving program fidelity. On the other hand, the long GHZ path of the logical CX gate between distant logical qubits would affect the parallelism between logical CX gates since two parallel logical CX gates cannot have intersected GHZ paths. That is, higher logical connectivity may hurt the latency. For any input program, we can use the QEC-CS compiler to predict the overhead of running a specific program on QEC-CS architectures with different logical connectivity. With these data obtained, we can then tune the logical qubit layout design accordingly to achieve a lower error rate or space-time overhead for the considered program. 

\noindent\textbf{Observation 2}: The placement of logical qubits affects the performance of logical CX gates and the `logical' connectivity. By co-designing the logical qubit layout design and the compiler design, we may further improve the fidelity or reduce the latency of specific programs.

\subsection{Compiler design for QEC-CS}\label{sec:designcomp}
Different implementations of non-Clifford gates (e.g., T gate) often induce different designs for compiling quantum programs on QEC architectures. 
Unfortunately, existing QEC compilers cannot enable efficient quantum computing on QEC-CS. They do not support the optimization of code switching, which is the critical difference between QEC-CS to other QEC architectures. 
To reduce the amount of code switching required by a quantum program, we need to address two important problems. 
The first problem is determining the logical gates we expect to execute with code switching. From the perspective of Steane code, the nontransverse logical gate is $T_L$ and it should be executed by using code switching. From the perspective of the RM code, we should execute the nontransverse $H_L$ by using code switching. The second problem is determining the number of logical gates we expect to execute after one invocation of code switching. The Steane code and RM code share many transverse logical gates, e.g., $CX_L$, $X_L$. If we are going to perform a code switching from the Steane code mode to the RM code mode for $T_L$, it is unclear whether we should switch back to the Steane code directly after the $T_L$ executed or perform more transverse logical gates (e.g., $CX_L$) in the RM code before switching back.
To answer the first question, we argue that most logical gates should be executed in the Steane code mode and use code switching for $T_L$. The insight is that the logical gate of the Steane code is more reliable than the ones of the RM code since there are fewer error locations (e.g., gates in the error detection circuit) in the Steane code, though being of the same distance as the RM code~\cite{chamberland2017error}. Thus, we should avoid executing too many logical gates on the RM code mode of QEC-CS logical qubits.
While using code switching for $T_L$ is better than $H_L$, it does not mean it is optimal to execute only one $T_L$ gate between a pair of code switching operations (i.e., Steane$\to$RM$\to$Steane). As shown in Figure~\ref{fig:codeconv}, the code switching operation contains a Steane logical CX gate, three physical gates between $q_{15}$ and other data qubits, and several ($\ge 5$) error detection rounds, making it far more time-consuming and error-prone than logical gates of the Steane code. 
Executing more logical gates in the RM code may increase error rates, but it may also reduce the count of code switching, thus improving fidelity, especially reducing latency. 
We observe that it may be more advantageous to execute more than one $T_L$ gate after one code switching operation, depending on the program context. 
To give an example where executing more than one $T_L$ after code switching may be more advantageous, let us consider the gate sequence in Equation~\ref{equ:gate}:
\begin{equation}\label{equ:gate}
    T_L\ q_0;T_L\ q_1;CX_L\ q_0\,q_1;T_L\ q_1,
\end{equation}
which is frequently appeared in various quantum programs~\cite{wille2008revlib} (e.g, arithmetic circuits, Grover). Two logical qubits ($\bar{q}_0$ and $\bar{q}_1$) are first switched from the Steane code to the RM code to perform the first two $T_L$ gates. If we switch the two logical qubits back to the Steane code after the two $T_L$ are executed, we would need another two code switching to perform the last $T_L$ on $\bar{q}_1$. 
However, if we delay the switching of two logical qubits behind the last $T_L\ \bar{q}_1$ executed, then we can save two code switching on $\bar{q}_1$. Such a delay of code switching is of merit, as long as the error rate/latency of two code switching is larger than the error rate/latency difference between the RM CX and the Steane CX. Indeed, this condition always holds as shown in Section~\ref{sect:eval}.
However, we are not going to delay all code switching since having too many logical gates executed in the RM code may hurt overall program fidelity as discussed before. We can achieve a good balance by inspecting the program context (e.g., the gate sequence in Equation~\ref{equ:gate}) and determining whether to delay code switching according to fidelity/latency gain.

\noindent\textbf{Observation 3}: Most logical gates should be executed in the Steane mode of QEC-CS logical qubits while the code switching should be applied in a context-aware way to promote the computational benefit of the QEC-CS architecture.

\section{Our Framework}
In this section, we introduce the detailed implementation of the QEC-CS architecture and compiler based on  considerations and observations outlined in Section~\ref{sec:design}.
\subsection{The QEC-CS architecture}
\begin{figure*}[h]
    \centering
    \includegraphics[width=0.98\textwidth]{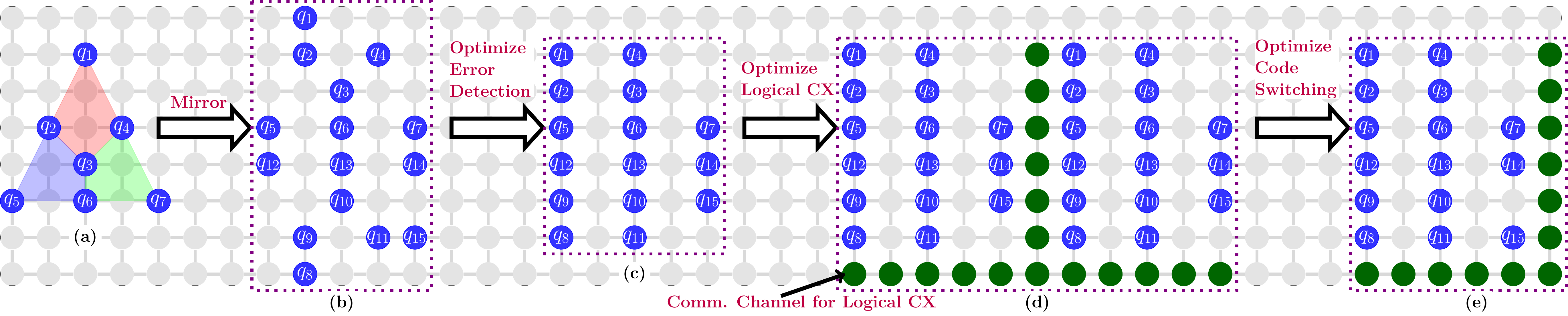}\hphantom{-8pt}
    \caption{Optimizing the data qubit layout of one QEC-CS logical qubit with knowledge from error detection, code switching, and the logical CX.}
    \label{fig:optimizelogicalqubit}\hphantom{-8pt}
\end{figure*}
\begin{figure}[h]   
    \includegraphics[width=0.48\textwidth]{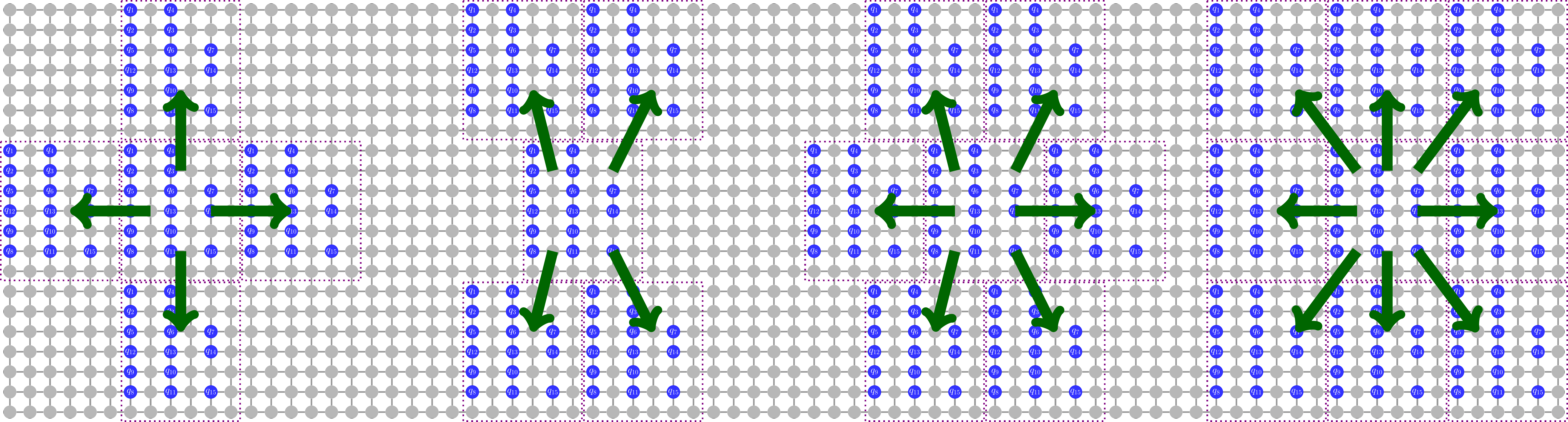}\hphantom{-4pt}
    
    {\scriptsize\hphantom{} \hspace{20pt} (a) \hspace{51pt} (b) \hspace{51pt} (c) \hspace{51pt} (d)
    \hphantom{-8pt}}
    \caption{Examples of placing multiple logical qubits. Green arrows denote logical CX directions. (a) connectivity 4. (b) connectivity 4 rotated. (c) connectivity 6. (d) connectivity 8.}
    \label{fig:optimizemultiqubit}\hphantom{-8pt}
\end{figure}
\subsubsection{Implementing one QEC-CS logical qubit}\hphantom{-11pt}

In this section, we search for the data qubit layout of one QEC-CS logical qubit according to \textbf{Observation 1}. 
The first step of the search is to guarantee minimal overhead of error detection circuits.
When implementing the error detection circuit of any QEC code, to guarantee full-distance error detection/correction~\cite{Chao2019FlagFE}, it is inevitable to use the flag-bridge circuit shown in Figure~\ref{fig:stabilizercirc}. Besides the parity qubit, we need to use at least one flag qubit for the weight-four stabilizer in the Steane code and RM code, and at least three flag qubits for the weight-eight stabilizer in the RM code. 
Under the requirement, the goal of this optimization step is to reduce the physical gate overhead of error detection circuits since more gates in error detection circuits indicate more error locations. Because of the potential connectivity limitation of the underlying  hardware, we may use ancillary qubits to help build the error detection circuits~\cite{Lao2020FaulttolerantQE}. In this section, we use the search of the QEC-CS logical qubit based on the Steane code and RM code as an example. Our strategy can be extended to find the layout for general 2D color codes~\cite{CostofUniversality}.
We first search for the physical qubit layout of the Steane code mode of a QEC-CS logical qubit. Since the RM code is almost doubling the Steane code (see Figure~\ref{fig:codeconv}), determining the layout of the Steane code mode could simplify the successive search for the RM code mode. The insight of searching for the data qubit layout is to use the geometrical shape of stabilizer operators.
As shown in Figure~\ref{fig:codeconv}, there are spatial relations between data qubits of the Steane code, e.g., $q_{2}$ is on the bottom left of $q_{1}$. With those spatial relations, we are able to find an initial data qubit layout for the Steane code by placing data qubits according to the spatial relation, as shown in Figure~\ref{fig:optimizelogicalqubit}(a). 
With the data qubit layout of the Steane code mode found, the next step is to determine the layout of the RM code mode. An initial layout of the RM code mode can be found by mirroring the layout of the Steane code mode, as shown in Figure~\ref{fig:optimizelogicalqubit}(b). This initialization can reduce the overhead of some error detection circuits, e.g., the one that measures $Z_5Z_6Z_{12}Z_{13}$. We further fine-tune the data qubit layout for smaller overhead of error detection circuits according to Equation~\ref{equ:ec}:
\begin{equation}\label{equ:ec}\small
    \forall s\in S, \min_{ts\in \{\uparrow, \downarrow, \leftarrow, \rightarrow, nop\}} TOT\_CX(\{ts_i(q_i),\cdots\}),\quad \{q_i,\cdots\}\text{: qb of }s
\end{equation}
The function $TOT\_CX$ is computed as the total edge count of the smallest bridge tree that connects all data qubits in an error detection circuit $s$. The overall set of error detection circuits $S$ is organized by first weight-eight stabilizers and then weight-four stabilizers. The optimization by  Equation~\ref{equ:ec} would be repeated several times (< 3) for a smaller overhead of error detection.
Figure~\ref{fig:optimizelogicalqubit}(c) shows the optimized layout of the RM code mode of the QEC-CS logical qubit.
Further, we should reduce the cost of the logical CX gate and the code switching operation. For these two operations, the source of overhead is the GHZ path length between non-neighboring data qubits.
The goal of this step is thus to reduce the total lengths of GHZ paths while still guaranteeing a small overhead of error detection.
The insight for this step is that to support remote CX between vertically distant data qubits, there should be a vertical line of unoccupied physical qubits as the communication channel (see Figure~\ref{fig:optimizelogicalqubit}(d)). Likewise, for horizontally distant data qubits, there should be a horizontal line of unoccupied physical qubits as the communication channel. Such communication channels can be guaranteed by trying to move each data qubit along a horizontal or vertical direction at most \textbf{one step}, according to Equation~\ref{equ:cx} (`EC': error detection circuits):
{
\hphantom{-10pt}
\small
\begin{align}\label{equ:cx}
\min_{ts\in \{\uparrow, \downarrow, \leftarrow, \rightarrow, nop\}}\sum_{i} GHZ\_LEN(ts_i(q_i),ts_i(q^h_i))+ GHZ\_LEN(ts_i(q_i),\nonumber\\ ts_i(q^v_i)) + TOT\_CX(Steane\ EC + RM\ EC)
\end{align}
\hphantom{-15pt}
}

\noindent The function $GHZ\_LEN$ is computed as the length of the shortest uninterrupted GHZ path between data qubits. $q^h_i$ and $q^v_i$ denote the qubits of the horizontal and vertical logical qubit neighbors, respectively. Figure~\ref{fig:optimizelogicalqubit}(d) shows the resulting layout for the QEC-CS logical qubit. This small tweaking of the data qubit layout will not greatly influence the overhead of error detection.
The optimization for the logical CX gate would also provide sufficient communication channels for the `interior' logical CX of the code switching operation. Thus, to reduce the overhead of remote physical CX in code switching, we would find the best location of $q_{15}$ according to Equation~\ref{equ:cs} (`CS': code switching):
\begin{equation}\label{equ:cs}
\small
    \min_{loc(q_{15})} TOT\_GHZ\_LEN(CS) + TOT\_CX(Steane\ EC + RM\ EC)
\end{equation}
Overall, the tuned layout of the QEC-CS logical qubit is shown in Figure~\ref{fig:optimizelogicalqubit}(e).

\subsubsection{Placing multiple QEC-CS logical qubits}\hphantom{-11pt}
\begin{table}[t]
  \centering
  \caption{Performance of logical CX gates for different logical qubit placement. The error rate of logical CX is computed when the device error rate is $10^{-6}$. The latency is normalized to the physical CX count. %
  }\label{tab:logicalcx}%
  \resizebox{0.46\textwidth}{!}{
\begin{tabular}{|l|l|c|c|c|c|}
\hline
\multicolumn{2}{|l|}{Logical CX gates} & \parbox{1cm}{Connect-ivity-4} & \parbox{2cm}{Connectivity-4 rotated} & \parbox{1cm}{Connect-ivity-6} & \parbox{1cm}{Connect-ivity-8} \\
\hline
\parbox{1.5cm}{\vspace{3pt}Horizontal}  & Latency & 72.1  & ---    & 72.1  & 72.1 \\
\cline{2-6}          
\parbox{1.5cm}{\hphantom{-3pt}Steane CX} & Error Rate & $3.9*10^{-9}$ & ---    & $3.9*10^{-9}$ & $3.9*10^{-9}$ \\
\hline
\parbox{1.5cm}{\vspace{3pt}Vertical} & Latency & 20.6  & ---    & ---    & 20.6 \\
\cline{2-6}        
\parbox{1.5cm}{\hphantom{-3pt}Steane CX}  & Error Rate & $3.3*10^{-9}$ & ---    & ---    & $3.3*10^{-9}$ \\
\hline
\parbox{1.5cm}{\vspace{3pt}Diagonal} & Latency & ---    & 30.9  & 30.9  & 72.1 \\
\cline{2-6}          
\parbox{1.5cm}{\hphantom{-3pt}Steane CX} & Error Rate & ---    & $3.8*10^{-9}$ & $3.8*10^{-9}$ & $5.3*10^{-9}$ \\
\hline
\parbox{1.5cm}{\vspace{3pt}Horizontal} & Latency & 82.4  & ---    & 82.4  & 82.4 \\
\cline{2-6}          
\parbox{1.5cm}{\hphantom{-3pt}RM CX} & Error Rate & $3.7*10^{-8}$ & ---    & $3.7*10^{-8}$ & $3.7*10^{-8}$ \\
\hline
\parbox{1.5cm}{\vspace{3pt}Vertical} & Latency & 41.2  & ---    & ---    & 41.2 \\
\cline{2-6}         
\parbox{1.5cm}{\hphantom{-3pt}RM CX} & Error Rate & $2.7*10^{-8}$ & ---    & ---    & $2.7*10^{-8}$ \\
\hline
\parbox{1.5cm}{\vspace{3pt}Diagonal} & Latency & ---    & 51.5  & 51.5  & 154.5 \\
\cline{2-6}          
\parbox{1.5cm}{\hphantom{-3pt}RM CX} & Error Rate & ---    & $3.0*10^{-8}$ & $3.0*10^{-8}$ & $3.9*10^{-8}$ \\
\hline
\end{tabular}%
}\hphantom{-8pt}
\end{table}

There is a large amount of freedom when placing logical qubits. We can tile logical qubits along horizontal, vertical, or diagonal directions, as shown in Figure~\ref{fig:optimizemultiqubit}, where logical qubit layouts with different connectivity are demonstrated.  Table~\ref{tab:logicalcx} shows the latency and fidelity of logical CX gates on different layouts. The data are obtained by simulating the QEC code under the circuit noise model. 
Layouts with different connectivity have their own advantages. Higher-connectivity layouts usually induce fewer SWAP gates than lower-connectivity layouts, thus can produce more reliable program outcomes. For example, for the connectivity-4 layout, if we perform a (Steane) logical CX gate along the diagonal direction, we need one logical SWAP gate along the vertical direction plus one horizontal logical CX gate, leading to an error rate of $1.4*10^{-8}$ when the device error rate is $10^{-6}$ (see Table~\ref{tab:logicalcx}). In contrast, on the connectivity-8 layout, the diagonal (Steane) logical CX gate is directly executable, with a lower error rate at $5.3*10^{-9}$ (see Table~\ref{tab:logicalcx}). Thus, establishing direct logical CX gates between distant logical qubits to increase connectivity may boost the reliability of the resulting QEC-CS layout. 
However, higher connectivity of the layout may instead hurt the parallelism in logical CX gates. For two logical CX gates, if their communication channels (green dots in Figure~\ref{fig:optimizelogicalqubit}) intersect with each other, then these two logical CX gates cannot be executed at the same time. For example, on the connectivity-8 layout, for a logical qubit $q_i$, denoting its upper left, upper, left logical qubits as $q_{i+1}, q_{i+2}, q_{i+3}$, $CX_L q_i q_{i+1}$ and $CX_L q_{i+2} q_{i+3}$ cannot be concurrently executed, though these qubits are neighboring to each other in the layout. Generally, the larger distance of two logical qubits of one logical CX gate is, the more logical CX gates may lag behind. Moreover, the communication channels of data qubits in the same logical qubit may also interfere with each other. As shown in Table~\ref{tab:logicalcx}, in the connectivity-8 layout, the latency of diagonal logical CX gates is far longer than the vertical logical CX gates.
Different layouts may fit programs of different features and thus provide opportunities for architecture-compiler co-design. For example, if one compiled program does not have much parallelism between logical CX gates, e.g., the UCCSD benchmark, using the connectivity-8 layout may be better than other layouts in terms of fidelity and latency of executed logical CX gates. On the other hand, for the compiled program without many global and long-distance logical CX gates, using the connectivity-4 (rotated) layout may reduce the space-time overhead of the program while still maintaining the same level of fidelity. Therefore, we can pick the best layout for a compiled program according to its logical CX feature. This co-design can further promote the computational potential of fault-tolerant quantum computing based on code switching.
Overall, in Section~\ref{sect:eval}, we have provided more quantified performance data for qubit layouts in Figure~\ref{fig:optimizemultiqubit} and demonstrated the benefits of architecture-compiler co-design.

\subsection{Compiler for QEC-CS}
\begin{table}[t]
\centering
\caption{Operations of the logical qubit in Figure~\ref{fig:optimizelogicalqubit}(e). The normalization stays the same for $p_e < 10^{-4}$. EC: error detection and correction.}\label{table:gateop}
\renewcommand*{\arraystretch}{1.12}
\resizebox{0.48\textwidth}{!}{
\begin{tabular}{|c|c|c|c|c|c|}
\hline
Logical op & \multicolumn{1}{c|}{RM CX} & \multicolumn{1}{c|}{RM 1q gate} & \multicolumn{1}{c|}{Steane CX} & \multicolumn{1}{c|}{Steane 1q gate} & \multicolumn{1}{c|}{Code switching} \\ \hline
\lcell{1.6cm}{\centering Normalized\\Infidelity} & \lcell{1.6cm}{\centering$\sim$8.8\\ Steane CX} & \lcell{1.6cm}{\centering$\sim$2.6\\ Steane CX} & \lcell{1.6cm}{\centering1.0\\ Steane CX} & \lcell{1.6cm}{\centering$\sim$0.2\\ Steane CX} & \lcell{1.6cm}{\centering$\sim$4.1 Steane CX} \\ \hline
\lcell{1.6cm}{\centering Normalized\\Latency} & \lcell{1.6cm}{\centering$\sim$5.5\\ Steane EC} & \lcell{1.6cm}{\centering$\sim$3.0\\ Steane EC} & \lcell{1.6cm}{\centering$\sim$2.9\\ Steane EC} & \lcell{1.6cm}{\centering$\sim$1.0\\ Steane EC} & \lcell{1.6cm}{\centering$\sim$9.1\\ Steane EC} \\ \hline
\end{tabular}
}\hphantom{-8pt}
\end{table}
To provide program compilation support for the QEC-CS architecture, we need to address two major tasks: QEC-CS architecture abstraction and code switching optimization.

\paragraph{QEC-CS architecture abstraction.}
The first task is to abstract the QEC-CS architecture. For comprehensive compiler optimization, this abstraction should expose the logical qubit topology and basic operations (i.e., the ISA) of the QEC-CS architecture. The coupling graph of logical qubits can be directly extracted from the given QEC-CS architecture. 
As for the ISA of the QEC-CS architecture, we would expose two more instructions besides Clifford+T logical gates. 
The first instruction is \textbf{EC} which means to perform error detection and correction at the current time point. The instruction is helpful for quantum computing involving many logical qubits where the independence of error correction on each logical qubit is demanded. 
The second instruction is \textbf{CS} which means to perform the code switching process at the current time point. Since the resource and time overhead of switching Steane code to RM code and switching RM code to Steane code is almost the same, one instruction for the code switching process is enough. The \textbf{CS} instruction provides the compiler with the ability to control code switching and unveil the computational advantage of the QEC-CS architecture, as discussed in Section~\ref{sec:design}.
Further, to achieve a fine-grained compilation, it is important to obtain the fidelity and latency data of each instruction of the QEC-CS architecture. The fidelity of the CS instruction and logical gates (e.g., $CX_L$, $H_L$, $T_L$) is simulated by our QEC simulator that adopts the circuit noise model, the stabilizer tableau representation~\cite{gidney2021stim} and the lookup table decoder. Note that when simulating the fidelity of the logical gate, we assume there is one EC instruction following the logical gate. For the QEC-CS architecture, the fidelity of the logical CX gate may not be uniform since the GHZ path length of the logical CX gate may vary. We would label each edge in the coupling graph of logical qubits with the fidelity of the logical CX related to the edge. The quantification of the non-uniformity of logical CX gates provides the compiler with the ability to perform fidelity-aware  routing.
As for the instruction latency, the time overhead of the CS and EC operations can be directly estimated from their specific implementations in the QEC-CS architecture. The latency of single-logical-qubit gates is almost equivalent to the latency of one physical single-qubit gate. The latency of logical CX contains two parts: the GHZ state preparation, and the remote CX gate protocol based on the GHZ state, as shown in Figure~\ref{fig:ghz}. Since the preparation of the GHZ state is to perform two rounds of CX gates between neighboring physical qubits, the latency of GHZ preparation is not related to the GHZ path length. Thus, all logical CX gates have the same latency data and can be estimated from Figure~\ref{fig:ghz}.

Overall, with the proposed abstraction of the QEC-CS architecture, we can adapt existing Clifford+T compilers to provide initial optimizations (e.g., reducing T gate counts, and communication routing) for quantum programs on the QEC-CS architecture. However, existing compilers are far from optimal for the QEC-CS architecture since they do not optimize the code switching operation which is critical for the QEC-CS architecture.
\paragraph{Optimization of Code Switching.} The second task is to reduce the usage of the code switching which is necessary but has a negative impact on the fidelity and time overhead of compiled programs. Conventionally, we would use two code switching operations for a logical T gate with one before it and one behind it. Here, we propose a novel compiler pass that reduces the amount of the code switching operations required by a quantum program in a context- and fidelity-aware way, as discussed in Section~\ref{sec:designcomp} (\textbf{Observation 3}). We describe our compiler pass for code switching optimization with the following steps.

\noindent\textbf{Step 1: Blocking.} For a given circuit, we search for the next (logical) T gate. Assuming the program qubit associated with this T gate is $q_0$, we create a new block $blk$ that contains the T gate. We denote the current qubit list of  $blk$ by $qlist$. For $q_i$ in $qlist$, we would also add other gates that are close to $blk$ and are applied on $q_i$ to $blk$  until the new gate on $q_i$ is H gate. We will update $qlist$ in this process since CX gates may have been added to $blk$. We would continue to add new gates to $blk$ until $qlist$ is not changed. This block represents the largest scope of the code switching on $q_0$.

\noindent\textbf{Step 2: Gate reordering.} In this step, will try to move the CX, X, and Z gates of $blk$ outside this block by circuit writing. Note that the $X$ and $Z$ can always be moved out with commuting rules in~\cite{Nam2017AutomatedOO}. This step is to remove unnecessary gates outside the scope of the code switching on $q_0$. In this way, we can keep most gates executed in the Steane code mode of QEC-CS logical qubits, and only execute necessary gates in the RM code mode. This step can reduce the latency and improve the fidelity of executing $blk$ with code switching.

\noindent\textbf{Step 3: Block refining.} For each CX gate $g$ in $blk$, computing the fidelity of $blk$ if we mark $g$ to be executed by the Steane code mode. If the fidelity is improved, we would split $blk$ into two parts, each part representing a new scope of the code switching on $q_0$. We will repeat this step until no splitting is possible. The goal of this step is to find the largest scope for the code switching on $q_0$ where no fidelity loss is caused by executing logical gates in the RM code mode.
We will repeat the above three steps until no new blocks are found. Finally, we would add the CS instructions at the start and end of each refined block. The start and end of each refined block in Step 3 represent the optimized timing of applying code switching on related logical qubits. 
In many cases, the refined block in Step 3 contains more than one logical T gate, which enables more efficient utilization of code switching.

\begin{figure}[h]
    \centering
    \includegraphics[width=0.48\textwidth]{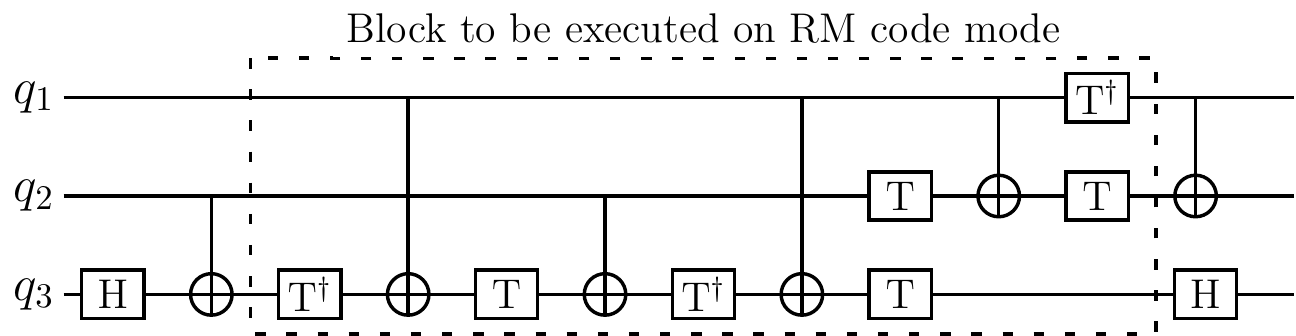}\hphantom{-5pt}
    \caption{An circuit (decomposed Toffoli gate) for illustrating code switching optimization.}
    \label{fig:examcirc}\hphantom{-5pt}
\end{figure}

As an example of the proposed code switching optimization, let us consider the circuit in Figure~\ref{fig:examcirc}. Without code switching optimization, for the logical qubit in Table~\ref{table:gateop}, the code switching count, infidelity (normalized to \# Steane CX), and latency (normalized to \# Steane EC) for the example circuit are 14, 82.0, and 167.8, respectively. With the proposed code switching optimization, the block found is highlighted with dotted lines in Figure~\ref{fig:examcirc}. Then, the code switching count, infidelity, and latency for the example circuit are reduced to 6 (by 57.1\%), 80.4 (by 2.0\%), and 105.4 (by 37.2\%), respectively.

\section{Evaluation}\label{sect:eval}
In this section, we first evaluate the logical qubit design and then evaluate the optimization of code switching as well as the co-design between architecture and compiler.

\subsection{Experiment Setup}\label{sect:expset}
\paragraph{Benchmark programs} We consider two categories of benchmark programs obtained from~\cite{li2022qasmbench}, as shown in Table~\ref{tab:benchmark}. The first category of benchmarks focuses on implementing arithmetic functions. These quantum programs are subroutines of large quantum applications. The second category of benchmarks aims to solve practical problems, e.g., Grover's algorithm, quantum walking, and Unitary Coupled Cluster ansatzes (UCCSD). 
For quantum walking, we specifically select the Binary Welded Tree (BWT) algorithm. For UCCSD, we simulate the $\text{CH}_4$ molecule. All programs are decomposed into the Clifford+T gate basis.

\renewcommand{\underline}[1]{#1}

\begin{figure}[t]
    \centering
    \includegraphics[width=0.48\textwidth]{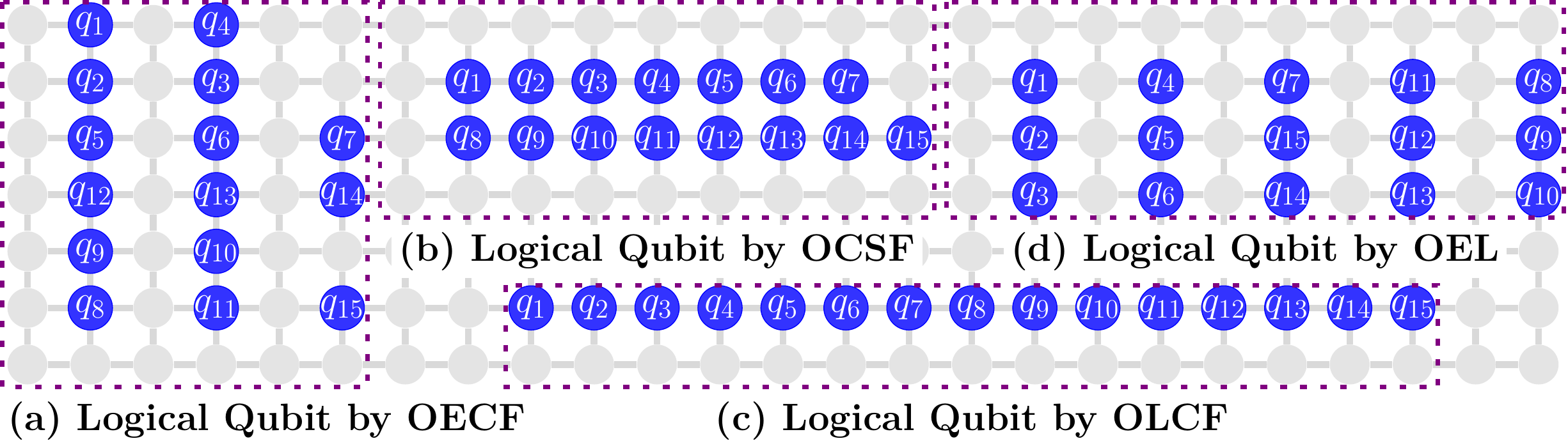}\hphantom{-6pt}
    \caption{QEC-CS logical qubits by four different schemes. }
    \label{fig:logicalqubit}\hphantom{-8pt}
\end{figure}

\paragraph{Baseline} We are the first full-stack framework for QEC-CS. Different baselines here are designed to  unveil the huge space of architecture and compiler optimization and provide an in-depth analysis of our design.
For architecture design, four schemes for generating QEC-CS logical qubits are evaluated. The first one, named \textit{OECF}, \underline{o}ptimizes the resource overhead of \underline{e}rror detection \underline{c}ircuits as the \underline{f}irst priority. OECF is the layout design proposed in this paper. The second one, named \textit{OCSF}, \underline{o}ptimizes the resource overhead of the \underline{c}ode \underline{s}witching operation as the \underline{f}irst priority. The third one, named \textit{OLCF}, \underline{o}ptimizes the resource overhead of the \underline{l}ogical \underline{C}X gate as the \underline{f}irst priority. The fourth one, named \textit{OEL},  \underline{e}venly places data qubits of a logical qubit respecting the underlying architecture. It first allocates the location of data qubits and then permutes the location of these data qubits for the best error correction capability, cheapest logical CX gates, and code switching in turn.
Figure~\ref{fig:logicalqubit} shows logical qubit layouts generated by the four schemes.
For compiler optimizations for code switching, we evaluate two schemes. The first one, named \textit{AgnosticCS}, is context-agnostic which always uses two code switching operations for a logical T gate (Steane mode $\to$ RM mode $\to$ Steane mode). AgnosticCS is the common strategy of using code switching in existing works. The second one, named \textit{AwareCS}, is context-aware which executes usually more than one logical gate in the RM mode, and would always try to reduce the usage of code switching as long as no fidelity loss is observed. AwareCS is the optimization proposed in this paper. For compiler optimizations except for code switching, they are not the focus of this paper and we adopt existing compiler toolkits for them. Specifically, we use PyZX~\cite{kissinger2019pyzx} to optimize T gate counts and Qiskit~\cite{Qiskit} to optimize CX gate counts (from unitary decomposition and remote CX gate routing).

\paragraph{Metircs} For the architecture design, the main metrics are the resource overhead, latency and fidelity of error detection, code switching, and logical CX gates. These metrics quantify the performance of one QEC-CS logical qubit. Besides, we also consider the space-time overhead (`\# physical qubit used * latency') and SWAP cost required to execute benchmark programs, in order to evaluate the performance of the placement of multiple logical qubits. 
For the compiler design, three major metrics are considered. The first one is the code switching count  required for a quantum program. We expect to reduce code switching operations. The remaining two are the overall program fidelity and latency, which are conventional compiler metrics.  %
\paragraph{Device noise model and error decoding} 
We assume a circuit noise model~\cite{gidney2021stim} with a physical error probability $p_e$ for the single-qubit depolarizing error channel on single-qubit gates, the two-qubit depolarizing error channel on two-qubit gates, and the Pauli-X error channel on measurement and reset operations. 
For error decoding of the QEC-CS architecture, we use the look-up table decoder for correcting data qubit errors induced by one physical gate. We would add one error detection round behind each logical gate.
\begin{table*}[t]
\caption{Resource, latency, and fidelity data of QEC-CS logical qubit layouts. OECF, OCSF, OLCF: optimize the  resource overhead of error detection circuits (EC), code switching (CS), logical CX in the first priority, respectively. OEL: optimize even distributed data qubit layout.
`OECF' is the one proposed in this paper. The operation latency data are normalized to physical CX counts. REM-CX: physical CX gate between non-neighboring qubits. }\label{table:archlayout}
\centering
\resizebox{0.99\textwidth}{!}{
\begin{tabular}{|l|llllll|lllll|lllll|}
\hline
\multirow{2}{*}{\lcell{1cm}{Arch Layouts}} & \multicolumn{6}{c|}{Resource Overhead} & \multicolumn{5}{c|}{Operation Latency} & \multicolumn{5}{c|}{Pseudo-threshold} \\ \cline{2-17} 
  & \multicolumn{1}{l|}{\lcell{1cm}{Tot. \# CX for Steane EC}} & \multicolumn{1}{l|}{\lcell{1cm}{Tot. \# CX for RM EC}} & \multicolumn{1}{l|}{\lcell{1.4cm}{Avg. REM-CX length for CS}} & \multicolumn{1}{l|}{\lcell{1.6cm}{Avg. REM-CX length for Steane $CX_L$}} & \multicolumn{1}{l|}{\lcell{1.4cm}{Avg. REM-CX length for RM $CX_L$}} & \lcell{1.2cm}{\# physi- cal qubit per logi- cal qubit} & \multicolumn{1}{l|}{\lcell{0.85cm}{Steane EC gate}} & \multicolumn{1}{l|}{\lcell{0.7cm}{RM EC gate}} & \multicolumn{1}{l|}{\lcell{1.cm}{Code Switching}} & \multicolumn{1}{l|}{\lcell{0.4cm}{Avg. Steane $CX_L$}} & \lcell{0.4cm}{Avg. RM $CX_L$} & \multicolumn{1}{l|}{\lcell{1cm}{Steane 1q gate}} & \multicolumn{1}{l|}{\lcell{1.2cm}{Avg. Steane $CX_L$}} & \multicolumn{1}{l|}{\lcell{1cm}{RM 1q gate}} & \multicolumn{1}{l|}{\lcell{1.1cm}{Avg. RM $CX_L$}} & \lcell{1cm}{Code switching} \\ \hline
\textbf{OECF} & \multicolumn{1}{l|}{40} & \multicolumn{1}{l|}{152} & \multicolumn{1}{l|}{3.7} & \multicolumn{1}{l|}{9.64} & \multicolumn{1}{l|}{10.73} & 42 & \multicolumn{1}{l|}{24.8} & \multicolumn{1}{l|}{74.4} & \multicolumn{1}{l|}{225.1} & \multicolumn{1}{l|}{46.35} & 61.80 & \multicolumn{1}{l|}{$1.4*10^{-3}$} & \multicolumn{1}{l|}{$2.8*10^{-4}$} & \multicolumn{1}{l|}{$1.1*10^{-4}$} & \multicolumn{1}{l|}{$3.1*10^{-5}$} & $6.8*10^{-5}$ \\ \hline
OCSF & \multicolumn{1}{l|}{68} & \multicolumn{1}{l|}{268} & \multicolumn{1}{l|}{1.7} & \multicolumn{1}{l|}{11.14} & \multicolumn{1}{l|}{10.83} & 36 & \multicolumn{1}{l|}{69.2} & \multicolumn{1}{l|}{187.8} & \multicolumn{1}{l|}{487.2} & \multicolumn{1}{l|}{41.20} & 82.40 & \multicolumn{1}{l|}{$5.5*10^{-4}$} & \multicolumn{1}{l|}{$1.5*10^{-4}$} & \multicolumn{1}{l|}{$3.8*10^{-5}$} & \multicolumn{1}{l|}{$1.6*10^{-5}$} & $2.9*10^{-5}$  \\ \hline
OLCF & \multicolumn{1}{l|}{68} & \multicolumn{1}{l|}{236} & \multicolumn{1}{l|}{7.5} & \multicolumn{1}{l|}{9.50} & \multicolumn{1}{l|}{9.50} & 30 & \multicolumn{1}{l|}{69.2} & \multicolumn{1}{l|}{190.2} & \multicolumn{1}{l|}{529.8} & \multicolumn{1}{l|}{30.9} & 56.65 & \multicolumn{1}{l|}{$5.5*10^{-4}$} & \multicolumn{1}{l|}{$1.7*10^{-4}$} & \multicolumn{1}{l|}{$4.8*10^{-5}$} & \multicolumn{1}{l|}{$2.1*10^{-5}$} & $2.5*10^{-5}$  \\ \hline
OEL & \multicolumn{1}{l|}{40} & \multicolumn{1}{l|}{212} & \multicolumn{1}{l|}{6.7} & \multicolumn{1}{l|}{9.57} & \multicolumn{1}{l|}{9.67} & 40 & \multicolumn{1}{l|}{24.8} & \multicolumn{1}{l|}{115.4} & \multicolumn{1}{l|}{277.4} & \multicolumn{1}{l|}{36.05} & 56.65 & \multicolumn{1}{l|}{$1.4*10^{-3}$} & \multicolumn{1}{l|}{$2.8*10^{-4}$} & \multicolumn{1}{l|}{$5.9*10^{-5}$} & \multicolumn{1}{l|}{$2.4*10^{-5}$} & $3.8*10^{-5}$  \\ \hline
\end{tabular}
}\vspace{-2pt}
\end{table*}

\begin{table*}[t]
\centering
\caption{Compilation results on the square grid of logical qubits generated by OECF. `Normalized infidelity': program infidelity normalized to Steane logical CX counts (for $p_e < 10^{-4}$). `Fidelity point': program fidelity  when $p_e = 10^{-6}$. The latency data is normalized to physical CX counts.}\label{tab:benchmark}
\resizebox{0.99\textwidth}{!}{
\begin{tabular}{|ll|l|l|l|lllll|lllll|}
\hline
\multicolumn{2}{|l|}{\multirow{2}{*}{Program}} & \multirow{2}{*}{\# qubit} & \multirow{2}{*}{\# gate} & \multirow{2}{*}{\# CX} & \multicolumn{5}{c|}{OECF+AgnosticCS} & \multicolumn{5}{c|}{\textbf{OECF+AwareCS} (Optimizations proposed in this paper)} \\ \cline{6-15} 
\multicolumn{2}{|l|}{} &  &  &  & \multicolumn{1}{l|}{\lcell{1.6cm}{Normalized infidelity}} & \multicolumn{1}{l|}{\lcell{1cm}{Fidelity point}} & \multicolumn{1}{l|}{Latency} & \multicolumn{1}{l|}{\lcell{1.6cm}{Space-time Overhead}} & \# CS & \multicolumn{1}{l|}{\lcell{1.6cm}{Normalized infidelity}} & \multicolumn{1}{l|}{\lcell{1cm}{Fidelity point}} & \multicolumn{1}{l|}{Latency} & \multicolumn{1}{l|}{\lcell{1.6cm}{Space-time Overhead}} & \# CS \\ \hline
\multicolumn{1}{|l|}{\multirow{3}{*}{\lcell{1.6cm}{Elementary Function}}} & \lcell{2cm}{Multi-qubit XOR} & 20000 & $2.19*10^6$ & $1.79*10^6$ & \multicolumn{1}{l|}{$4.76*10^6$} & \multicolumn{1}{l|}{98.3\%} & \multicolumn{1}{l|}{$1.47*10^8$} & \multicolumn{1}{l|}{$1.24*10^{14}$} & $5.60*10^5$ & \multicolumn{1}{l|}{$4.72*10^6$} & \multicolumn{1}{l|}{98.3\%} & \multicolumn{1}{l|}{$1.01*10^8$} & \multicolumn{1}{l|}{$8.51*10^{13}$} & $3.20*10^5$ \\ \cline{2-15} 
\multicolumn{1}{|l|}{} & \lcell{2cm}{Ripple-Carry Adder} & 30000 & $2.27*10^6$ & $1.97*10^6$ & \multicolumn{1}{l|}{$4.20*10^6$} & \multicolumn{1}{l|}{98.5\%} & \multicolumn{1}{l|}{$9.85*10^7$} & \multicolumn{1}{l|}{$1.24*10^{14}$} & $4.20*10^5$ & \multicolumn{1}{l|}{$4.17*10^6$} & \multicolumn{1}{l|}{98.6\%} & \multicolumn{1}{l|}{$6.39*10^7$} & \multicolumn{1}{l|}{$8.35*10^{13}$} & $2.40*10^5$ \\ \cline{2-15} 
\multicolumn{1}{|l|}{} & \lcell{2cm}{Ripple-Carry Comparator} & 30000 & $2.39*10^6$ & $2.06*10^6$ & \multicolumn{1}{l|}{$4.29*10^6$} & \multicolumn{1}{l|}{98.4\%} & \multicolumn{1}{l|}{$1.00*10^8$} & \multicolumn{1}{l|}{$1.27*10^{14}$} & $4.20*10^5$ & \multicolumn{1}{l|}{$4.26*10^6$} & \multicolumn{1}{l|}{98.5\%} & \multicolumn{1}{l|}{$6.63*10^7$} & \multicolumn{1}{l|}{$8.35*10^{13}$} & $2.40*10^5$ \\ \hline
\multicolumn{1}{|l|}{\multirow{3}{*}{\lcell{1.6cm}{Quantum Application}}} & Grover & 38 & $1.05*10^6$ & $6.30*10^5$ & \multicolumn{1}{l|}{$3.71*10^6$} & \multicolumn{1}{l|}{98.7\%} & \multicolumn{1}{l|}{$1.33*10^8$} & \multicolumn{1}{l|}{$2.12*10^{11}$} & $5.79*10^5$ & \multicolumn{1}{l|}{$3.66*10^6$} & \multicolumn{1}{l|}{98.7\%} & \multicolumn{1}{l|}{$8.56*10^7$} & \multicolumn{1}{l|}{$1.37*10^{11}$} & $3.31*10^5$ \\ \cline{2-15} 
\multicolumn{1}{|l|}{} & BWT Oracle & 28 & $1.03*10^6$ & $5.98*10^5$ & \multicolumn{1}{l|}{$3.81*10^6$} & \multicolumn{1}{l|}{98.6\%} & \multicolumn{1}{l|}{$1.38*10^8$} & \multicolumn{1}{l|}{$1.62*10^{11}$} & $6.06*10^5$ & \multicolumn{1}{l|}{$3.77*10^6$} & \multicolumn{1}{l|}{98.6\%} & \multicolumn{1}{l|}{$8.80*10^7$} & \multicolumn{1}{l|}{$1.03*10^{11}$} & $3.46*10^5$ \\ \cline{2-15} 
\multicolumn{1}{|l|}{} & UCCSD & 16 & $1.07*10^6$ & $1.79*10^5$ & \multicolumn{1}{l|}{$4.70*10^6$} & \multicolumn{1}{l|}{98.3\%} & \multicolumn{1}{l|}{$2.36*10^8$} & \multicolumn{1}{l|}{$1.58*10^{11}$} & $8.23*10^5$ & \multicolumn{1}{l|}{$4.70*10^6$} & \multicolumn{1}{l|}{98.3\%} & \multicolumn{1}{l|}{$2.36*10^8$} & \multicolumn{1}{l|}{$1.58*10^{11}$} & $8.23*10^5$ \\ \hline
\end{tabular}
}\vspace{-8pt}
\end{table*}
\begin{figure*}[h]
    \includegraphics[width=0.99\textwidth]{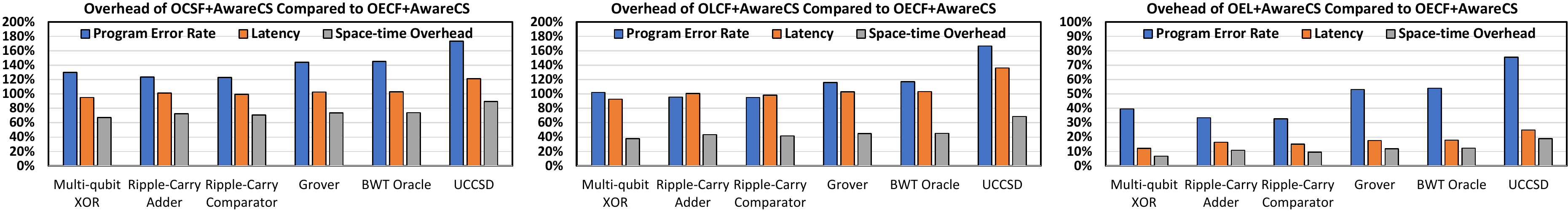}\hphantom{-4pt}
    
    {\scriptsize\hphantom{}\hspace{85pt}(a)\hspace{160pt}(b)\hspace{160pt}(c)}
    \hphantom{-8pt}
    \caption{Data in the plot is `overhead of the tested logical qubit design$/$overhead of OECF'*100\%-1. Logical qubits are arranged in a square grid.}
    \label{fig:archeffect}\hphantom{-5pt}
\end{figure*}
\begin{figure*}[h!]
    \includegraphics[width=0.99\textwidth]{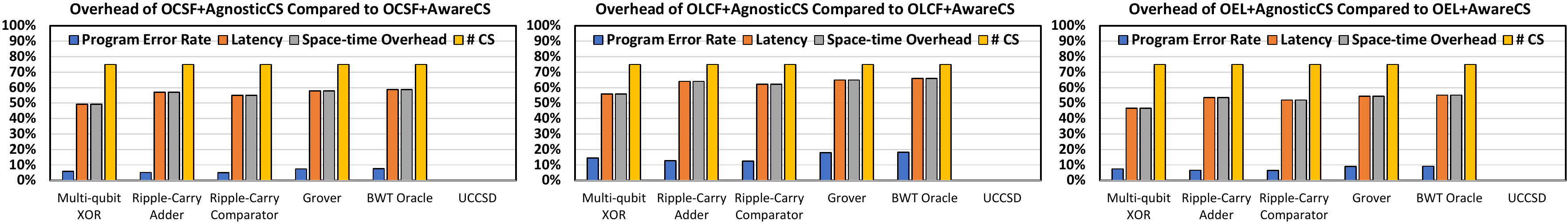}\hphantom{-4pt}
    
    {\scriptsize\hphantom{}\hspace{85pt}(a)\hspace{160pt}(b)\hspace{160pt}(c)}
    \hphantom{-8pt}
    \caption{
    Data in the plot is (`overhead of AgnosticCS$/$overhead of AwareCS'*100\%-1)
     on logical qubits generated by OCSF, OLCF and OEL.}
    \label{fig:conveffect}\hphantom{-5pt}
\end{figure*}
\begin{figure*}[h!]
    \includegraphics[width=0.99\textwidth]{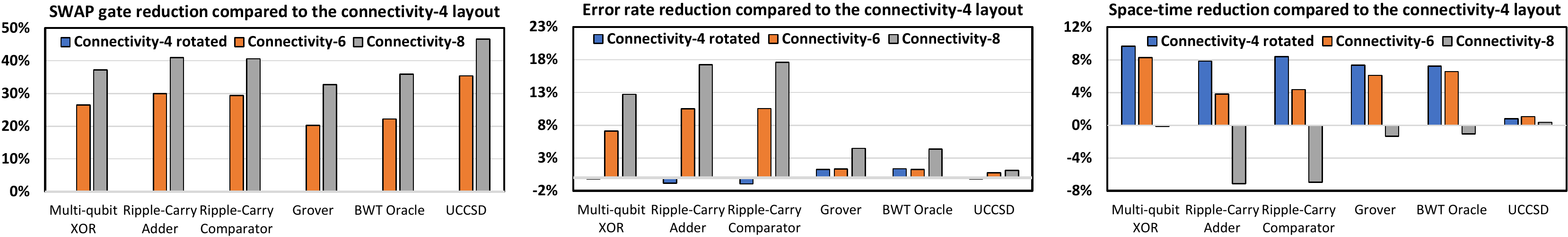}\hphantom{-4pt}
    {\scriptsize\hphantom{}\hspace{85pt}(a)\hspace{160pt}(b)\hspace{160pt}(c)}
    \hphantom{-8pt}
    \caption{The effect of connectivity when placing multiple logical qubits generated by OECF.}
    \label{fig:archcomp}\hphantom{-12pt}
\end{figure*}

\subsection{Experiment Results}

Table~\ref{table:archlayout} shows the performance of the different logical qubit designs in terms of resource overhead, latency, and fidelity. The `REM-CX length' in Column 4-6 refers to the Manhattan distance between two data qubits we are going to perform a physical CX gate on. The `REM-CX length' is equal to the GHZ path length plus one.
Figure~\ref{fig:logicalqubit} shows the performance of different logical qubit designs in the quantum program context. Table~\ref{tab:benchmark} and Figure~\ref{fig:archeffect} illustrate the performance of code switching optimizations proposed in the paper. Figure~\ref{fig:archcomp} demonstrates the effect of connectivity when placing multiple logical qubits. `Connectivity-4/6/8' means to let each logical qubit have 4/6/8 neighboring logical qubits when placing multiple logical qubits. The `Connectivity-4 rotated' architecture is achieved by rotating the `Connectivity-4' architecture by $\frac{\pi}{4}$, as shown in Figure~\ref{fig:optimizemultiqubit}.
Overall, compared to other logical qubit designs (OCSF, OLCF, and OEL), the proposed logical qubit design (OECF) on average reduces the error rate, space-time overhead, and latency of test programs by 99.3\%, 43.9\% and 75.1\%, respectively. Moreover, on the proposed logical qubit design, compared to AgnosticCS, the proposed AwareCS reduces the space-time overhead and code switching counts of test programs by 43.8\% and 62.5\%, respectively. This benefit of AwareCS is not limited by the logical qubit design. Finally, the results show that increasing the connectivity between logical qubits does not necessarily induce a better FTQC platform and different programs  favor different logical qubit layouts. We elaborate on these conclusions in the following analysis.

\hphantom{-5pt}\subsubsection{The effect of logical qubit design}\hphantom{-10pt}

\paragraph{Firstly, it is critical to reduce the resource overhead of error detection circuits as the first priority when designing a logical qubit.} As shown in Table~\ref{table:archlayout}, compared to OCSF and OLCF, OECF has higher resource overhead for code switching and logical CX gates, respectively. However, OECF has the highest error threshold for all logical operations. This indicates that error detection has the largest impact on the fidelity of logical operations and OECF's low overhead for error detection guarantees the reliability of logical operations. The OEL logical qubit has the second-best error threshold, also due to its specific optimizations for error detection circuits. Moreover, OECF logical qubits are the most reliable for quantum programs. As shown in Figure~\ref{fig:logicalqubit}, OECF reduces the error rate of (six) test programs on average by 138.8\%, 113.1\%, and 46.0\%, compared to OCSF, OLCF, and OEL, respectively.

\paragraph{Secondly, the latency of error detection circuits is also the most important factor in the space-time overhead of quantum programs.} Though the OECF logical qubit has the most physical qubits per logical qubit (see Table~\ref{table:archlayout} Column 6), the smallest latency overhead of OECF logical qubits (see Table~\ref{table:archlayout} Column 2, 3) still guarantees the smallest space-time overhead for quantum programs. As shown in Figure~\ref{fig:logicalqubit}, OECF reduces the space-time overhead of (six) test programs on average by 74.4\%, 46.0\%, 11.2\%, compared to OCSF, OLCF, and OEL, respectively. This is because each logical gate of a quantum program is often followed by an error detection operation. This amplifies the effect of the error detection latency for the overall program latency. Indeed, as shown in Figure~\ref{fig:logicalqubit}, OECS reduces the latency of test programs on average by 103.4\%, 104.8\%, 17.0\%, compared to OCSF, OLCF, and OEL, respectively.

\paragraph{Thirdly, the optimization of logical CX gates is more important than the optimization of code switching.} With similar error detection overhead, OCSF and OLCF show better code switching and logical CX gates, due to their specific optimizations toward code switching and logical CX gates, respectively. Inspecting Figure~\ref{fig:logicalqubit}(a)(b), OLCF induces 9.2\% lower program error rate and 15.8\% lower space-time overhead, on average for test programs. This is because, after the proposed compiler optimization for code switching, the number of logical CX gates is on average 3.38 times more than the number of code switching (see Table~\ref{tab:benchmark} Column 5, 15) in test programs, making the optimization of logical CX gates more advantageous. 
Ideally, for most Toffoli-gate based quantum programs, without considering the overhead of SWAP-based routing, the logical T gate count is at most 16.7\% larger than the logical CX gate count (see Figure~\ref{fig:examcirc}). On the other hand, as shown in Table~\ref{tab:benchmark}, one logical T gate on average induces 1.14 code switching operations (see Table~\ref{tab:benchmark} Column 10 and 15). This means in a Toffoli-gate based program, the number of code switching operations is at most 33.0\% larger than the number of logical CX gates. Thus, the logical CX count is larger than the code switching count in the routed circuit, as long as each logical CX gate requires 0.11 logical SWAP gates for routing. Therefore, in most connectivity-constrained architectures, optimizing logical CX gates is more critical than optimizing code switching. 

\paragraph{Finally, incorporating QEC code information into the layout design is critical for logical qubit design.} Compared to OECF, OEL first determines the data qubit layout according to the underlying hardware topology. Though we have tried to improve the performance of the logical qubit by OEL (in terms of error detection, logical CX and code switching), the logical operations by OECF, especially the ones related to the RM code mode, are more reliable than those by OEL, as shown in Table~\ref{table:archlayout} Column 11-15. Moreover, as shown in Figure~\ref{fig:logicalqubit}, compared to OECF, OEL on average increases the program error rate, latency, and space-time overhead of test programs by 46.0\%, 17.0\% and 11.2\%, respectively.

\hphantom{-5pt}\subsubsection{The effect of code switching optimization}\hphantom{-10pt}

The optimization of code switching is critical for reducing the space-time overhead of quantum programs. As shown in Table~\ref{tab:benchmark}, compared to `OECF+AgnosticCS', `OECF+AwareCS' on average reduces the space-time overhead, and code switching count of test programs by 43.8\% and 62.5\%, respectively. This demonstrates that the code switching optimization would induce significant space-time overhead reduction. This observation can be further explained with two facts. Firstly, code switching widely exists in test programs. For quantum programs executed with `OECF+AgnosticCS', the code switching count is on average about 62.2\% of the logical CX count. Secondly, the code switching operation is far more time-consuming than logical gates. As shown in Table~\ref{table:archlayout}, the latency of code switching is 8.72, 2.99, 3.16, and 1.65 times the latency of `Steane 1q gate+EC', `Steane CX gate+EC', `RM 1q gate+EC', and `RM CX gate+EC', respectively. As for the fact that AwareCS does not show benefits on the UCCSD benchmark, it is because each logical T gate in the UCCSD circuit is followed by a logical H gate, providing no opportunities for code switching optimization. Fortunately, this is not a common pattern in quantum programs, especially for Toffoli-gate-based programs (e.g., Grover).
Further, the benefit of AwareCS is significant even when logical qubit design changes. As shown in Figure~\ref{fig:conveffect}, compared to AgnosticCS, AwareCS on average reduces the space-time overhead of test programs on OCSF, OLCF, and OEL logical qubits by 46.4\%, 52.1\%, and 43.6\%, respectively. This is because the logical qubit design will not change the occurrence of code switching in quantum programs and the latency of code switching is always far longer than other logical operations in different logical qubit designs (see Table~\ref{fig:logicalqubit} Column 8-12).
The benefit of AwareCS would become even more remarkable when the logical qubit connectivity is higher. This is because the logical CX count of test programs on highly-connected architecture would be smaller, amplifying the effect of code switching. 

\hphantom{-5pt}\subsubsection{Architecture-Compiler Co-design}\hphantom{-10pt}

\paragraph{No layout is universally better:}
For physical quantum hardware with constrained connectivity between physical qubits, enforcing higher connectivity between logical qubits when placing multiple logical qubits does not necessarily induce a better computing platform. 
As shown in Figure~\ref{fig:archcomp}(b), compared to the connectivity-4 setting (where logical qubits form a square grid), the connectivity-6 and connectivity-8 setting does improve the fidelity of quantum programs because they greatly reduce the SWAP gate count as shown in Figure~\ref{fig:archcomp}(a). However, the connectivity-8 setting induces higher space-time overhead than the connectivity-4 setting, as shown in Figure~\ref{fig:archcomp}(c). This is because logical CX gates in the diagonal direction hurt the parallelism in logical CX gates and induce extra latency (see Table~\ref{tab:logicalcx}).
Also, as shown in Figure~\ref{fig:archcomp}(c), the connectivity-6 setting induces higher space-time overhead than the rotated connectivity-4 setting. In this case, the horizontal logical CX gate takes longer time than the diagonal logical CX gate (see Table~\ref{tab:logicalcx}). 
Overall, it is reasonable to match the connectivity of logical qubits with the connectivity of physical qubits. Enforcing higher connectivity between logical qubits may not simultaneously provide benefits to fidelity and space-time overhead.

\paragraph{Architecture-Compiler Co-design:} While it is not possible to achieve better performance of all programs by simply enforcing higher connectivity between logical qubits, it is possible to promote the performance of some specific quantum programs with the compiler output in Figure~\ref{fig:archcomp}. 
For example, to achieve the smallest space-time overhead while allowing slightly ($< 2\%$) higher error rate of test programs (than on the connectivity-4 layout), we can use the connectivity-6 layout for the UCCSD benchmark and `Connectivity-4 rotated' layout for the remaining programs. Likewise, to achieve the lowest error rate while allowing slightly ($< 2\%$) higher space-time overhead of test programs (than on the connectivity-4 layout), we can use the connectivity-6 layout for the ripple-carry adder and comparator benchmark and connectivity-8 layout for the remaining programs. 
The ability of adjusting QEC-CS architecture provides us the opportunity to co-design with the compiler to improve the performance of specific quantum programs.

\section{Related Work} 

\textbf{Compilers for Programs upon QEC codes:}
Lao et al.~\cite{Lao2018MappingOL}  proposed a mapping process to execute lattice surgery-based quantum circuits on surface code architectures.
Ding et al.~\cite{Ding2018MagicStateFU} and Paler et al.~\cite{Paler2019SurfBraidAC} studied the compilation of magic state distillation circuits, which is another way to implement the logical T gate.
Javadi et al.~\cite{JavadiAbhari2017OptimizedSC} and Hua et al.~\cite{Hua2021AutoBraidAF} studied the topological routing of logical CX gates over the surface code.
These works assume a readily available FTQC architecture and do not consider implementing the QEC code on hardware. Besides, those papers do not consider the optimization of the code switching operation. 

\noindent\textbf{Implementing the QEC architecture:}
Reichardt~\cite{Reichardt2018FaulttolerantQE} and Chamberland et al.~\cite{Chamberland2019TriangularCC} respectively proposed planar and trivalent qubit layouts to synthesize the color code.
Chamberland et al.~\cite{Chamberland2020TopologicalAS} introduced heavy architectures to map the subsystem code and the surface code. 
Wu et al.~\cite{surfstitch} proposed an automated method to stitch the surface code to superconducting quantum hardware. Those efforts focus on one QEC code and one logical qubit. Those works cannot be simply extended to support the code switching operation which involves dynamic conversion between two QEC codes, let alone providing the compiler optimization for code switching.

\section{Conclusion}

In this paper, we propose the first full-stack framework, named \frameworkName, from architecture design to compiler design, to enable fault-tolerant quantum computing based on code switching.
Our framework searches for architecture designs of logical qubits by inspecting the feature of QEC protocols, e.g., error detection and code switching. Afterward, our framework proposes context-aware compilation optimizations to avoid unnecessary invocations of the code switching operation, which are more erroneous and time-consuming than logical gates. Finally, our framework presents the archi-tecture-compiler co-designs to further unveil the computational potential of quantum computing based on code switching. Overall, our paper presents the first systematic exploration of code switching for FTQC, unveiling and crystallizing the extensive design space of QEC-CS. 

\section*{Acknowledgment}
This material is based upon work supported by the U.S. Department of Energy, Office of Science, National Quantum Information Science Research Centers, Quantum Science Center. Andrew W. Cross would like to acknowledge the support from the U.S. Department of Energy, Office of Science, National Quantum Information Science Research Centers, Co-design Center for Quantum Advantage (C2QA) under contract number DESC0012704. The Pacific Northwest National Laboratory is operated by Battelle for the U.S. Department of Energy under Contract DE-AC05-76RL01830. This work was also supported in part by NSF 2048144, NSF 2138437 and Robert N. Noyce Trust.

\bibliographystyle{unsrt}
\bibliography{references}
\end{document}